\documentclass[12pt]{iopart}

\usepackage{iopams}
\usepackage{setstack}
\usepackage{graphicx}
\usepackage{bm}
\usepackage{verbatim}
\usepackage{epstopdf}
\usepackage{tensor}
\usepackage{cite}
\DeclareMathAlphabet{\mathpzc}{OT1}{pzc}{m}{it}

\begin{document}


\title[Scalar and spinor quasinormal modes of an $f(R)$ global monopole]{Scalar and spinor quasinormal modes of an $f(R)$ global monopole}

\author{J. P. Morais Gra\c ca$^{1}$, H. S. Vieira$^{1}$ and V. B. Bezerra$^{1}$}

\address{$^{1}$ Departamento de F\'{i}sica, Universidade Federal da Para\'{i}ba, Caixa Postal 5008, CEP 58051-970, Jo\~{a}o Pessoa, PB, Brazil}

\ead{jpmorais@gmail.com, horacio.santana.vieira@hotmail.com and valdir@fisica.ufpb.br}

\begin{abstract}
The quasinormal modes for scalar and spinor fields in the background spacetime corresponding to an $f(R)$ global monopole are calculated using the WKB approximation. In the obtained results we emphasize the role played by the parameter $\psi_{0}$, associated with the $f(R)$ gravity. We discuss the appropriate limit $\psi_{0} \rightarrow 0$, in which case the results concerning to the Schwarzschild black hole are obtained, as it should be expected.
\end{abstract}




\maketitle


%
%
\section{Introduction}
The investigations concerning the interaction of black holes with matter and fields around then give us the possibility to get some information about the physics of these objects. One of these information can be obtained from the characteristic complex frequencies called quasinormal modes (QNMs) which arises from the perturbation in the state of the black hole induced by its interaction with matter and fields (For a review on QNMs of black holes, see \cite{Kokkotas:1999bd, Konoplya:2011qq}). The spectra of frequencies which are associated with damped oscillations depend only on the structure of the black hole spacetimes characterized by the mass, eletric and magnetic charges and angular momentum \cite{Schutz:1985zz}, do not depending on the initial perturbation. Thus, we can use the observational data \cite{Iyer:1986np} to estimate the values of the parameters associated with black holes and identify the existence of these objects, as well as we can use the result to get some information about their physics, from the classical point of view, or eventually, on their quantum properties. 

On the other hand, these modes are important to study the processes involving gravitational radiation by black holes, their stability and detection \cite{Nollert:1999ji}. Along this line of research a lot of investigations have been done in different contexts which include gravitational black holes \cite{Leaver:1990zz,Andersson:1996xw,Wang:2000gsa,Hod:2003jn,Berti:2003zu,Jing:2005pk,Yoshida:2010zzb,Yang:2012he,Cvetic:2013lfa,Richartz:2014jla}, as well as their acoustic analogues \cite{Oliveira:2015vqa,Okuzumi:2007hf,Barcelo:2007ru}.

Since the discovery that the universe is currently in an accelerated expansion era, several theories have been proposed to explain such an unexpected fact. One of them is the so-called $f(R)$ theory of gravity, where the Einstein-Hilbert action is replaced by a more general action involving a generic function of the Ricci scalar (For a review, see \cite{Sotiriou:2008rp}). In a cosmological context, the $f(R)$ theory introduces the possibility to explain both the late time cosmic speed up and the early time inflationary scenario in one same model, without the introduction of an $ad$ $hoc$ cosmological constant \cite{Nojiri:2009kx}. Also, $f(R)$ theory can be seen as an effective theory that introduces corrections to Einstein's gravity. To give an example, the Starobinsky model \cite{Starobinsky:1980te}, given by $f(R) = R + \alpha R^2$, is a model that nicely fits the latest results on Cosmic Microwave Background data obtained by the Planck satellite \cite{Planck:2013jfk}.

Topological defects were expected to be formed by a vacuum phase transition in the early universe. These defects include domain walls, cosmic strings and monopoles. The monopole as a topological defect was first introduced by t'Hooft \cite{'tHooft:1974qc} as a regular field configuration whose stability is guaranteed by its vacuum topological properties. It is composed by a self-coupled Higgs triplet $\phi^a$ and a non-abelian gauge field $A^a_\mu$ whose original local $O(3)$ symmetry is broken to $U(1)$. This remaining $U(1)$ symmetry is then responsible for a magnetic-like field whose behaviour can represent a point-like magnetic monopole. A global monopole can be seen as a similar kind of topological defect, but now with only a scalar triplet field under a global $O(3)$ symmetry and no gauge field. The gravitational properties of the global monopole was first investigated by Barriola and Vilenkin in the context of general relativity \cite{Barriola:1989hx}, and later extended to alternative theories of gravity like Brans-Dicke theory \cite{Barros:1997fi} and $f(R)$ theory \cite{Carames:2011uu}. 

Our proposal in this work is to study the stability of a global monopole inside a black hole in $f(R)$ theory of gravity. The thermodynamics of such system has been studied in \cite{Man:2013sf} and compared with the same system in Einstein's gravity. Also, the motion of classical massive particles was analyzed in this background spacetime \cite{Carames:2011xi}.

Our main goal is to calculate the quasinormal modes induced by external scalar and spinor field perturbations. In our context, quasinormal modes are resonant nonradial deformations of the black hole with a global monopole. These perturbations form a spectra of discrete and complex frequencies, where the real part determines the frequency of oscillation and the imaginary part determines the rate of damping of each mode due to the emission of radiation. 

In this paper we will calculate the scalar and spin $1/2$ modes in the mentioned background and compare our results with the ones obtained for the Schwarzschild solution. There are quite a few methods to evaluate quasinormal modes, and we have chosen to use a 3th order WKB schema as presented in \cite{Iyer:1986np}. For low-lying modes this schema fits well with numerical results, as show in \cite{Iyer:1986nq}. This paper is organized as follows: In section 2 we introduce the spacetime. In section 3 we introduce the scalar equation we must deal with together with a brief introduction of the WKB schema. After that we calculate the scalar quasinormal modes. In section 4 we introduce the Dirac equation in the considered spacetime and use the same WKB schema to calculate the spin $1/2$ quasinormal modes. In section 5, we present our conclusions.

\section{Global monopole spacetime in $f(R)$ gravity}
The metric generated by a static and spherically symmetric black hole with a global monopole in $f(R)$ gravity is given by \cite{Carames:2011uu}
\begin{equation}
ds^{2}=\Delta\ dt^{2}-\Delta^{-1}\ dr^{2}-r^{2}(d\theta^{2}+\sin^2\theta\ d\phi^{2})\ ,
\label{eq:metric_spherically_monopole}
\end{equation}
where
\begin{equation}
\Delta=1-\frac{2M}{r}-8\pi\eta^{2}-\psi_{0}r\ ,
\label{eq:Delta_metric_spherically_monopole}
\end{equation}
with $\psi_0$ being a constant which measures the deviation of this solution that corresponds, in fact, to a global monopole in $f(R)$ gravity, from the global monopole in general relativity \cite{Barriola:1989hx}. The parameter $M$ above is a mass parameter and we will consider it as the total mass of the black hole with a global monopole inside it. The $\eta$ parameter is related to the scale of symmetry breaking, which is of the order of $10^{16}$ GeV for a typical Grand Unified Theory (GUT). For consistence we should have $\eta^2 \ll \psi_{0} \ll 1$, otherwise (\ref{eq:metric_spherically_monopole}) is not the valid metric for the system under consideration \cite{Carames:2011uu}. We are using units where $G \equiv c \equiv \hbar \equiv k_{B} \equiv 1$.

From Eq.~(\ref{eq:Delta_metric_spherically_monopole}), we have that the horizon surface equation is obtained from the condition
\begin{equation}
\Delta=(r-r_{+})(r-r_{-})=0\ .
\label{eq:surface_hor_spherically_monopole}
\end{equation}
The solutions of Eq.~(\ref{eq:surface_hor_spherically_monopole}) are
\begin{equation}
r_{\pm}=\frac{1-8 \pi  \eta ^2\pm\sqrt{\left(1-8 \pi  \eta ^2\right)^2-8 M \psi_{0} }}{2 \psi_{0} }\ ,
\label{eq:sol_surface_hor_spherically_monopole_1}
\end{equation}
and so we have two horizons for this metric. The interior horizon can be considered the black hole event horizon and goes to $2M$ when $\eta = 0$ and $\Psi_{0} \rightarrow 0$, while the exterior one is a kind of cosmological horizon and blows up for $\eta = 0 $ and $\psi_{0} \rightarrow 0$. This second result is due to the fact that we don't have this exterior horizon in the Schwarzschild spacetime and the above metric should reduce to the Schwarzschild metric in the appropriate regime, namely, as $\eta = \psi_{0} = 0$.

%
%
\section{Scalar quasinormal modes}
The more straightforward way to study perturbations in a black hole spacetime is to introduce fields around it and demand that the fields do not produce a backreaction on the background. For a massive scalar field, the perturbations are described by the Klein-Gordon equation in a curved spacetime, which is given by the following covariant form
\begin{equation}
\left[\frac{1}{\sqrt{-g}}\partial_{\sigma}(g^{\sigma\tau}\sqrt{-g}\partial_{\tau})+\mu_{0}^{2}\right]\Psi=0\ ,
\label{eq:Klein-Gordon}
\end{equation}
where $\mu_{0}$ is a parameter that is identified with the mass of the scalar field.  

Substituting Eq.~(\ref{eq:metric_spherically_monopole}) into Eq.~(\ref{eq:Klein-Gordon}), we obtain
\begin{equation}
\left[-\frac{r^{2}}{\Delta}\frac{\partial^{2}}{\partial t^{2}}+\frac{\partial}{\partial r}\left(r^{2}\Delta\frac{\partial}{\partial r}\right)-\mathbf{L}^{2}-\mu_{0}^{2}r^{2}\right]\Psi=0\ ,
\label{eq:mov_spherically_monopole}
\end{equation}
where $\mathbf{L}^{2}$ is the angular momentum operator. 

Due to the time independence and symmetry of the spacetime with respect to rotations, the solution of Eq.~(\ref{eq:mov_spherically_monopole}) can be written as
\begin{equation}
\Psi=\Psi(\mathbf{r},t)=R(r)Y_{l}^{m}(\theta,\phi)\mbox{e}^{-i\omega t}\ ,
\label{eq:ansatz_solution_spherically_monopole}
\end{equation}
where $Y_{l}^{m}(\theta,\phi)$ are the usual spherical harmonics. Substituting Eq.~(\ref{eq:ansatz_solution_spherically_monopole}) into Eq.~(\ref{eq:mov_spherically_monopole}), we find that
\begin{equation}
\frac{d}{dr}\left(r^{2}\Delta\frac{dR}{dr}\right)+\left(\frac{r^{2}\omega^{2}}{\Delta}-\lambda_{lm}-\mu_{0}^{2}r^{2}\right)R=0\ ,
\label{eq:mov_radial_spherically_monopole}
\end{equation}
where $\lambda_{lm}=l(l+1)$. 

The form of the radial equation given by Eq.(\ref{eq:mov_radial_spherically_monopole}) is not appropriate to study the QNMs. In order to find the ideal form, let us perform a coordinate transformation, defined by the following relation

\begin{equation}
dr_{*} = \frac{dr}{\Delta}.
\end{equation}
The coordinate $r_{*}$ maps the region between the horizons into the $(-\infty,+\infty)$ region and is called the tortoise coordinate. The radial Klein-Gordon equation then reads

\begin{equation}
-\frac{d^2R}{dr_{*}^2} + V(r) = \omega^2R,
\end{equation}
where

\begin{equation}
V(r) = \Delta (\frac{l(l+1)}{r^2} + \frac{\Delta'}{r}),
\end{equation}
with the $prime$ indicating derivative with respect to $r$. We will consider $\mu_{0}$ for the sake of simplicity.

\subsection{The WKB method}

The WKB method is a fairly well know technique used to obtain approximated solutions to the one-dimensional time independent Schr{\"o}edinger equation. In the gravitational arena, it was applied by Schutz and Will \cite{Schutz:1985zz} to the problem of scattering on black holes.

Here we will use the WKB method to compute the quasinormal modes $\omega$. We will use the third order approximation as given by \cite{Iyer:1986np}. This method has been extensively used in different black hole configurations and is accurate up to a few percents for both the real and imaginary parts for low-lying modes with $n < l$. The method was further extended up to the sixth order in \cite{Konoplya:2003ii}.

\begin{figure}
\centering
\begin{tabular}{@{}lc@{}}
\includegraphics[scale=0.4]{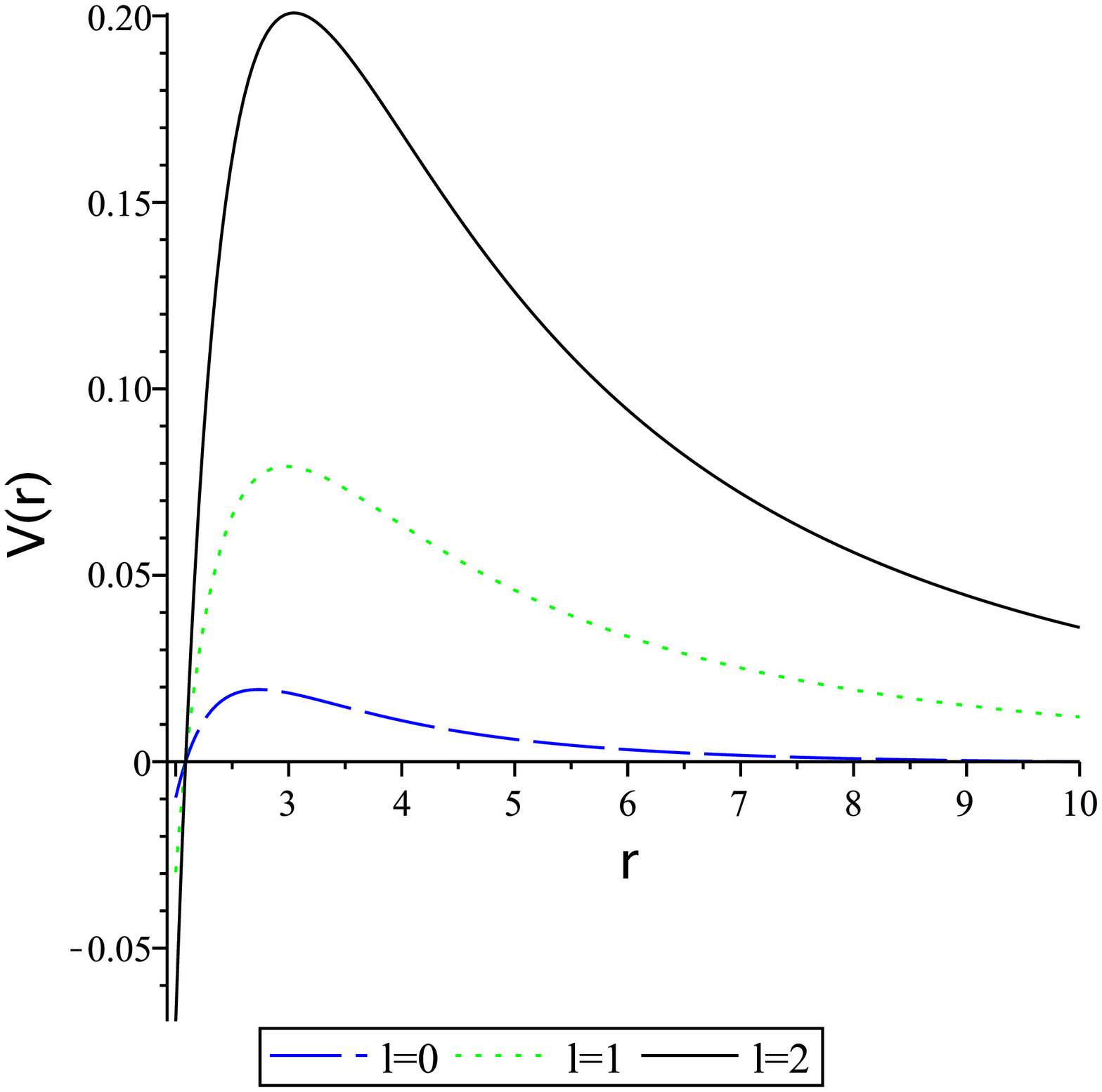} &
\includegraphics[scale=0.4]{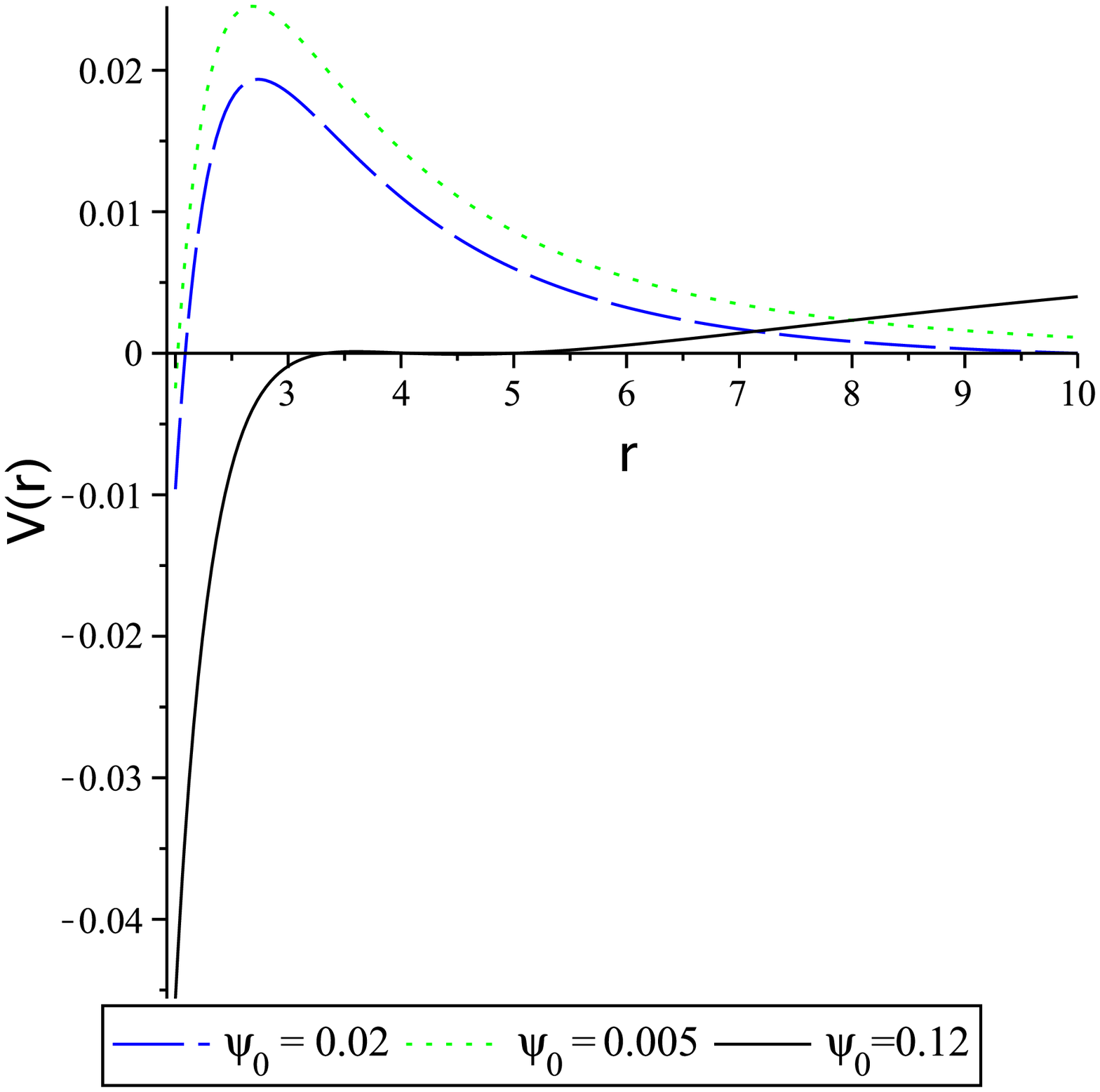} \\
\end{tabular}
\caption{On the left, the figure shows the effective V(r) potential for $8\pi\eta^2 \approx 0$ and $\psi_{0} = 0.02$, for several $l$ values. On the right, the effective potential for different $\psi_0$ values. We can see that for larger $\psi_0$ values we no longer have quasinormal modes, but in any case the above metric is only valid for $\psi_0 \ll 1$.}
\label{fig:effectivePotential}
\end{figure}
 
The formula for the complex quasinormal modes is given by

\begin{equation}
E^2 = [V_0 + (-2 V_0^{\prime\prime})^{1/2}\Lambda]  - i(n+\frac{1}{2})(-2V_0^{\prime\prime})^{1/2}(1+\Omega) 
\end{equation}
where

\begin{equation}
\Lambda = \frac{1}{(-2V_0^{\prime\prime})^{1/2}}[\frac{1}{8}(\frac{V_0^{(4)}}{V_0^{\prime\prime}})(\frac{1}{4} + \alpha^2) - \frac{1}{288}(\frac{V_0^{\prime\prime\prime}}{V_0^{\prime\prime}})^2(7+60\alpha^2)] ,
\end{equation}

\begin{eqnarray}
\Omega = \frac{1}{(-2V_0^{\prime\prime})}[\frac{5}{6912}(\frac{V_0^{\prime\prime\prime}}{V_0^{\prime\prime}})^4 (77 + 188 \alpha^2) - \frac{1}{384}(\frac{V_0^{\prime\prime\prime 2} V_0^{(4)}}{V_0^{\prime\prime 3}})(51 + 
\\
+ 100 \alpha^2) + \frac{1}{2304}(\frac{V_0^{(4)}}{V_0^{\prime\prime}})^2(67 + 68 \alpha^2) + \frac{1}{288}(\frac{V_0^{\prime\prime\prime}V_0^{(5)}}{V_0^{\prime\prime 2}})(19+28\alpha^2)
\\
- \frac{1}{288}(\frac{V_0^{(6)}}{V_0^{\prime\prime}})(5+4\alpha^2)] .
\end{eqnarray}
Also,

\begin{equation}
\alpha = n + \frac{1}{2} ,
\end{equation}

\begin{equation}
V_0^{(n)} = \frac{d^nV}{dr_{*}^n}\bigg|_{r_* = r_*(r_{max})} ,
\end{equation}
where $r_{max}$ stands for the radius where the effective potential presents its peak. The effective potential for the massless scalar field is given by

\begin{figure}[htb]
\centering
\begin{tabular}{@{}lc@{}}
\includegraphics[scale=0.3]{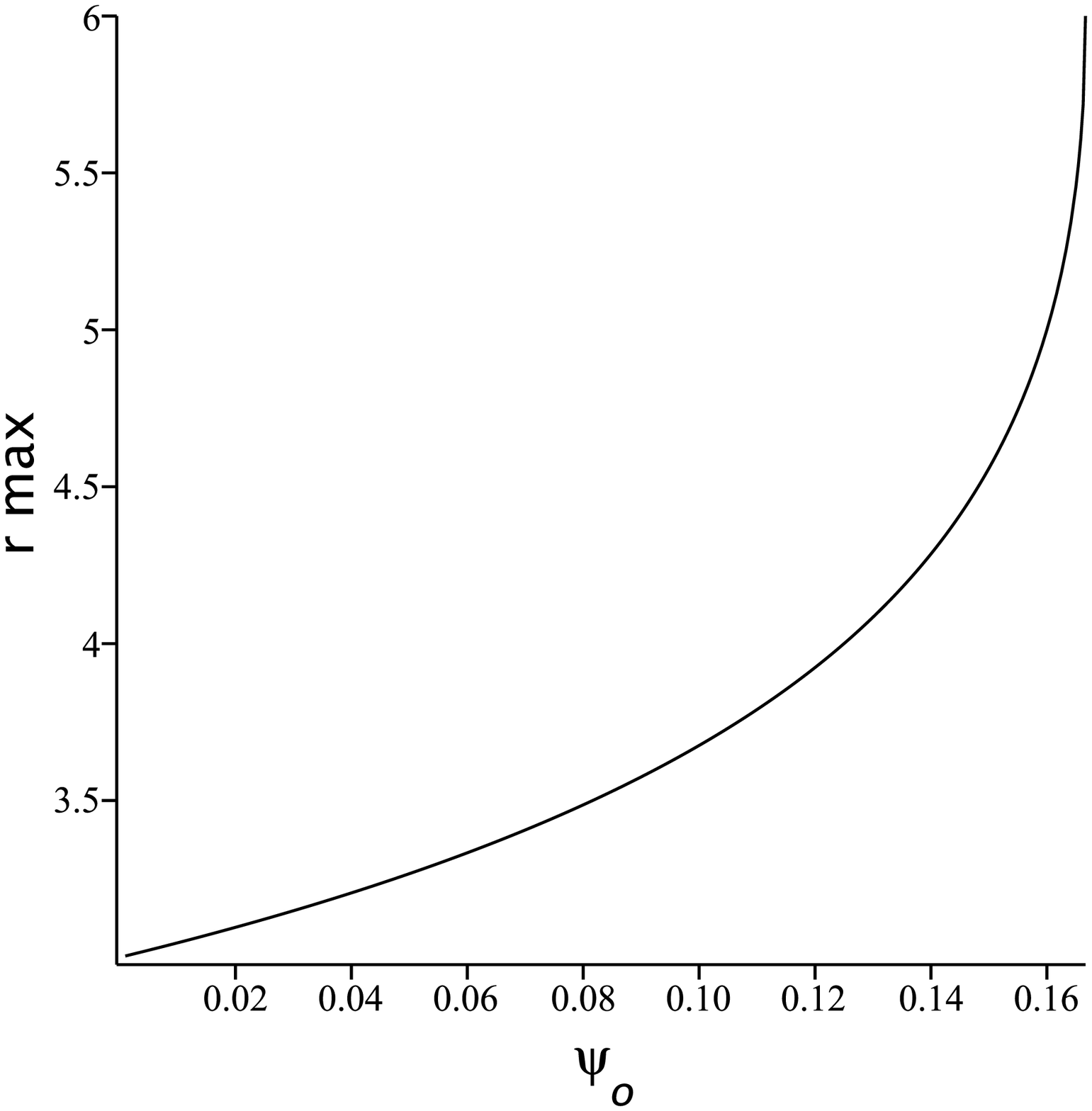} &
\includegraphics[scale=0.4]{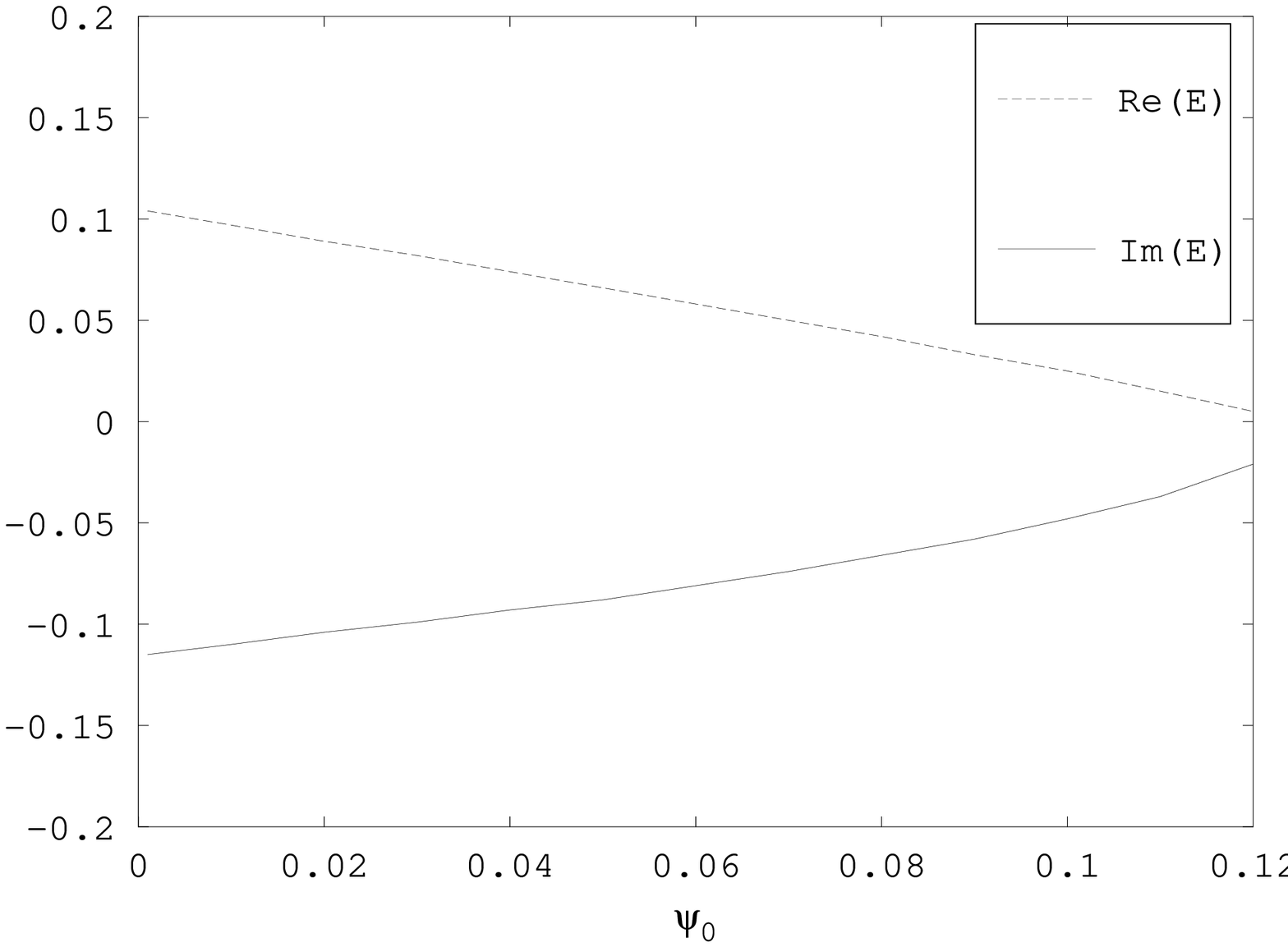} \\
\end{tabular}
\caption{The plot on the left shows the value of the radius where the effective potential reach its maximum. The plot on the right shows how the $n=0$, $l=0$ mode changes as we increase the parameter $\psi_{0}$. Both plots indicates that when the constraint $\psi_{0} \ll 1$ is no longer valid, all our schema appears to break apart.}
\label{fig:limits_Psi}
\end{figure}

\begin{equation}
V(r) = (a - \frac{2}{r} - \psi_{0}r)(\frac{l(l+1)}{r^2} + \frac{2}{r^3} - \frac{\psi_{0}}{r})
\end{equation}
where $a = 1 - 8\pi\eta^2$ and $M=1$. This effective potential resembles a barrier as it is shown in Figure (\ref{fig:effectivePotential}), for $a=1$ and $\psi_{0}=0.02$ for $l=0,1,2$. The position of the peak depends on $l$, and for the Schwarzschild black hole in Einstein's gravity we have $r_{max}(l\rightarrow \infty) \rightarrow 3$. In $f(R)$ gravity the peak will also depends on $\psi_{0}$ and $\eta$. For $l \rightarrow \infty$, and considering $8 \pi \eta^2 \approx 0$, we have the relation 

\begin{equation}
\psi_{0} r^2 - 2r + 6 = 0.
\end{equation}
In Figure (\ref{fig:limits_Psi}) we show a graph for the peak ($r_{max}$) as $l \rightarrow \infty$. We can see that as $\psi_{0}$ increases, $r_{max}$ also increases. For $\psi_{0} > 1/6$ the radius become complex, but as the given metric is only valid for $\psi_{0} \ll 1$, we don't need to worry about this regime. In the same Figure we also show how the $l=0,n=0$ mode changes as we increase $\psi_{0}$. In principle, the system appear to become unstable, since it will reach a region with non-negative values for the imaginary part. This is not actually true, because the WKB schema fails before we reach these values and even the metric is no longer valid as $\psi_{0} \rightarrow 1$.

\begin{figure}
\centering
\includegraphics[scale=0.7]{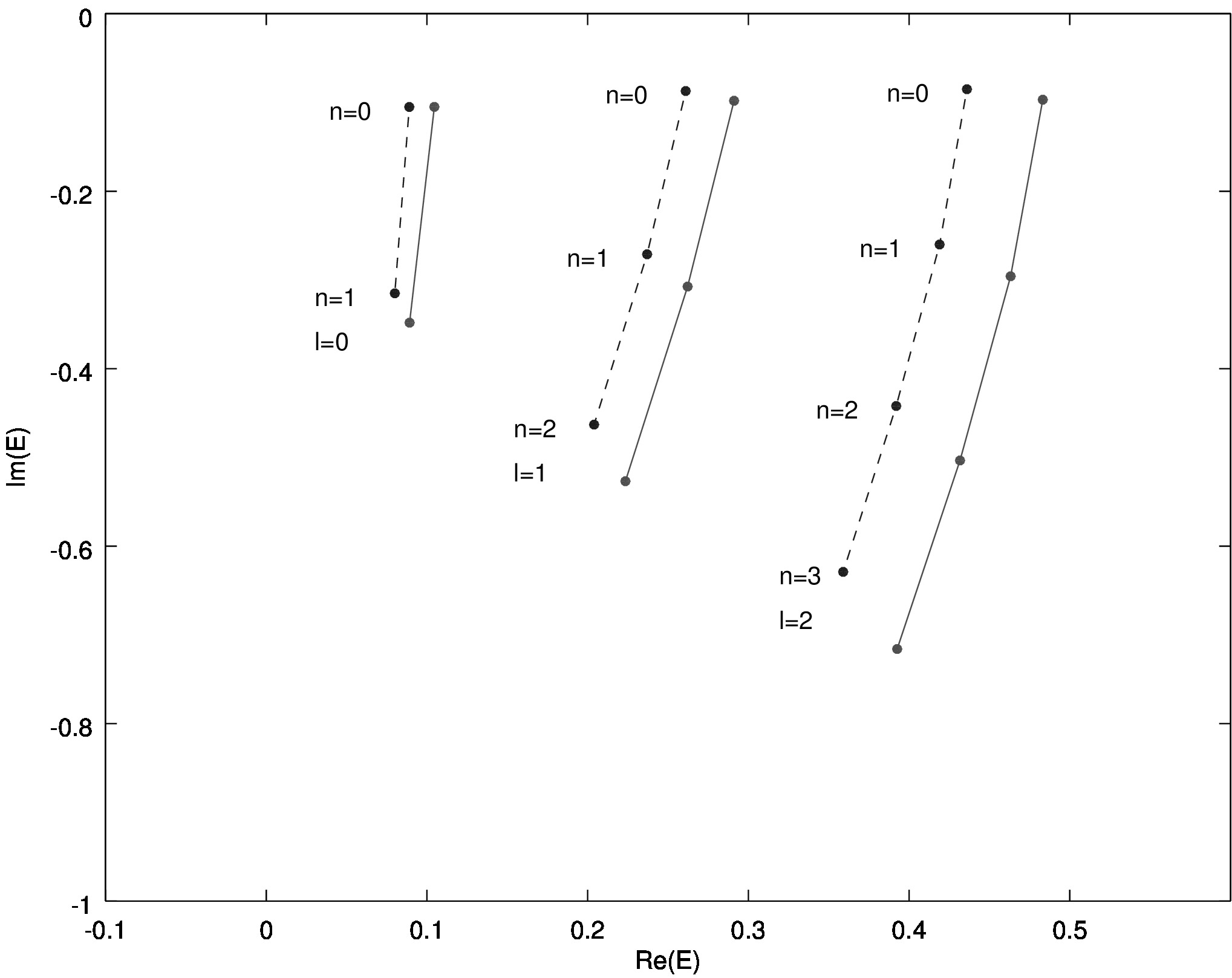}
\caption{The dotted line represents the massless scalar quasinormal modes for a black hole with a global monopole in $f(R)$ theory, for $\psi_{0} = 0.02$ and $8 \pi \eta^2 = 10^{-5}$. The continuous line represents the same modes for the Schwarzschild solution.}
\label{fig:qnm_scalar_comparasion}
\end{figure}

\subsection{Quasinormal modes for the massless scalar field}

The quasinormal modes for $n=0,1,2$ and $l \leq n+1$ are show in Table (\ref{tab:QNM_scalar}), for $\psi_{0} = 0.02$, $M = 1$ and $8 \pi \eta^2 = 10^{-5}$. Here we used the same values for the parameters that has been used in \cite{Man:2013sf}. In Figure (\ref{fig:qnm_scalar_comparasion}) we compare our results with the results obtained for the Schwarzschild black hole using the same schema. We can see that the inclusion of the parameter $\psi_{0}$ lower the values of the real part as well as the imaginary part of the modes, but does not affect its stability.  

\begin{table}[]
\centering
\caption{Scalar quasi-normal modes for $\psi_0 = 0.02$ and $8 \pi \eta^2 = 10^{-5}$.}
\label{tab:QNM_scalar}
\begin{tabular}{p{0.2\textwidth}p{0.2\textwidth}p{0.2\textwidth}p{0.2\textwidth}}
\hline
 l  & n  & Re(E) & -Im(E)  \\ \hline
 0  & 0  & 0.089 & 0.105  \\
 0  & 1  & 0.080 & 0.315  \\
 1  & 0  & 0.261 & 0.087  \\
 1  & 1  & 0.237 & 0.271  \\
 1  & 2  & 0.204 & 0.463  \\
 2  & 0  & 0.436 & 0.085  \\
 2  & 1  & 0.419 & 0.260  \\
 2  & 2  & 0.392 & 0.442  \\
 2  & 3  & 0.359 & 0.629  \\
\end{tabular}
\end{table}

\section{Spinor quasinormal modes}
The Dirac equation in a curved spacetime can be written using the vielbeins and the one-form spin connection. The vielbeins are used to define a local rest frame where we can introduce local Dirac matrices.

With this in mind, we can write the Dirac equation in a general background as

\begin{equation}
[\gamma^a e_a^{\;\mu}(\partial_\mu + \Gamma_\mu)+m]\Psi =  0
\label{eqn:Dirac}
\end{equation}
where $e_a^{\;\mu}$ are the vielbeins, a 4-set of 4-vectors defined by $e_a^{\;\mu} e_b^{\;\nu} g(x)_{\mu\nu} = \eta_{ab}$, the $\Gamma_\mu$ is the one-form spin connection which will be defined later and $\gamma^a$ are the flat spacetime Dirac matrices

\begin{equation}
\gamma^0 = \left( \begin{array}{cc}
-i & 0 \\
0 & i \end{array} \right), \hspace{10pt}
\gamma^i = \left( \begin{array}{cc}
0 & -i\sigma^i \\
i \sigma^i & 0 \end{array} \right).
\end{equation}  
As usual, $\sigma^i$ are the Pauli matrices. 
We will approach the above equation using the inverse of the previous defined vielbein, the one-forms $e_\mu^{\;a}$ defined by $e_\mu^{\;a} e_\nu^{\;b} \eta_{ab} = g(x)_{\mu\nu}$. These 4-set one-forms, that we will call tetrads to differentiate from their inverse, can be thought as a kind of square root of the metric and carry information about the local curvature of the spacetime. The way they differ from one point to another is given by their exterior derivative
\begin{equation}
d e^a \equiv - \omega^a_{\;b} \wedge e^b = - \omega^a_{\;b\mu} dx^\mu \wedge e^b_{\;\nu} dx^\nu
\end{equation}  
from which we can get $\Gamma_\mu$, by contract $\omega^a_{\;b}$ with the generators of the Lorentz transformations for spinors, that so $\Gamma_\mu = (1/4) \omega_\mu^{\;ab} \gamma_a \gamma_b$, where the Latin indices a,b,c... can be raised and lowered by the flat Minkowski metric $\eta_{ab}$.  

We take the tetrads to be
\begin{equation}
e_\mu^{\;a} = diag(\sqrt{\Delta}, \frac{1}{\sqrt{\Delta}},r,r sin\theta),
\end{equation}
and then the spin connections can be expressed as
\begin{eqnarray}
\Gamma_0 = \frac{1}{4} \Delta' \gamma_0 \gamma_1, \hspace{10pt} \Gamma_1 = 0, \hspace{10pt} \Gamma_2 = \frac{1}{2} \sqrt{\Delta}\gamma_1 \gamma_2
\nonumber
\\
\Gamma_3 = \frac{1}{2}(sin\theta \sqrt{\Delta} \gamma_1 \gamma_3 + \cos\theta \gamma_2 \gamma3) .
\end{eqnarray}
Thus, the Dirac equation given by (\ref{eqn:Dirac}), turns into
\begin{eqnarray}
-\frac{\gamma_0}{\sqrt{\Delta}}\frac{\partial \Psi}{\partial t} + \sqrt{\Delta} \gamma_1 (\frac{\partial}{\partial r} + \frac{1}{r} + \frac{1}{4 \Delta}\frac{\partial \Delta}{\partial r})\Psi +
\frac{\gamma_2}{r}(\frac{\partial}{\partial \theta} + \frac{1}{2} cot \theta)\Psi + 
\\
\nonumber
\frac{\gamma_3}{r sin\theta}\frac{\partial \Psi}{\partial \phi} + m \Psi = 0,
\end{eqnarray}
where $\Delta = 1 - \frac{2M}{r} - 8 \pi \eta^2 - \phi_0 r$. We can rescale $\Psi$ as $\Psi = \Delta^{-\frac{1}{4}}\Phi$ and get a simpler form for the Dirac equation, which can be written as

\begin{eqnarray}
-\frac{\gamma_0}{\sqrt{\Delta}}\frac{\partial \Phi}{\partial t} + \sqrt{\Delta} \gamma_1 (\frac{\partial}{\partial r} + \frac{1}{r})\Phi +
\frac{\gamma_2}{r}(\frac{\partial}{\partial \theta} + \frac{1}{2} cot \theta)\Phi + 
\\
\nonumber
\frac{\gamma_3}{r sin\theta}\frac{\partial \Phi}{\partial \phi} + m \Phi = 0.
\end{eqnarray} 

This equation is closely related to the spherically symmetric Dirac equation in a flat spacetime, and so we can write down

\begin{equation}
\Phi = \left( \begin{array}{c}
\frac{iG^{(\pm)}(r)}{r} \phi^{\pm}_{jm}(\theta,\psi) \\
\frac{F^{(\pm)}(r)}{r} \phi^{\mp}_{jm}(\theta,\psi) 
\end{array} \right) e^{-i\omega t},
\end{equation}  
where the spinor harmonics are given by

\begin{eqnarray}
\phi^{+}_{jm} = \left( \begin{array}{c}
\sqrt{\frac{j+m}{2j}} Y_l^{m-1/2} \\
\sqrt{\frac{j-m}{2j}} Y_l^{m+1/2}
\end{array} \right),\hspace{10pt} \left( \mbox{for j = l + } \frac{1}{2} \right),
\end{eqnarray}

\begin{eqnarray}
\phi^{-}_{jm} = \left( \begin{array}{c}
\sqrt{\frac{j+1-m}{2j+2}} Y_l^{m-1/2} \\
-\sqrt{\frac{j+1+m}{2j+2}} Y_l^{m+1/2}
\end{array} \right),\hspace{10pt} \left( \mbox{for j = l - } \frac{1}{2} \right),
\end{eqnarray}
with $Y^{m \pm 1/2}_{l}(\theta,\psi)$ being the usual spherical harmonics for spin-up and spin-down particles, respectively. We will repeat the approach used in the scalar case and redefine a radial coordinate transformation given by the relation

\begin{equation}
\frac{dr}{dr_{*}} = \Delta,
\end{equation}
in such a way to get an appropriate form of the radial Dirac equation to discuss the quasinormal modes. Doing this transformation, the Dirac equation can be written as

\begin{equation}
\hspace{-40pt}
\frac{d}{dr_{*}} \left( \begin{array}{c}
F^{\pm} \\ G^{\pm} \end{array} \right) - \Delta^{\frac{1}{2}} \left( \begin{array}{cc}
\kappa_{\pm}/r & m \\ m & -\kappa_{\pm}/r \end{array} \right) \left( \begin{array}{c}
F^{\pm} \\ G^{\pm} \end{array} \right) = \left( \begin{array}{cc}
0 & -\omega \\ \omega & 0 \end{array} \right) \left( \begin{array}{c}
F^{\pm} \\ G^{\pm} \end{array} \right) ,
\end{equation}
where $\kappa_{\pm}$ are the eigenvalues of the operator $\hat{K}^2 = \hat{J}^2 + 1/4$, and are given by

\begin{equation}
\kappa_{\pm} = \Bigg\{ \begin{array}{c} -(j+1/2), \hspace{20pt}j = l + 1/2 \\ (j+1/2),\hspace{20pt}j = l-1/2 \end{array}
\end{equation}
For the sake of simplicity, for now on we will consider the massless case. Also, it can be proved that the radial equations relating the two different sets of functions (plus and minus) are the same, and so the spin up and spin down cases equal. We will then simply call these functions $F$ and $G$. With this consideration, it is straightforward to decouple the two equations as

\begin{eqnarray}
\frac{d^2F}{dr_{*}^2} + (\omega^2 - V_1)F = 0
\label{eq:Dirac1}
\end{eqnarray}

\begin{eqnarray}
\frac{d^2G}{dr_{*}^2} + (\omega^2 - V_2)G = 0 ,
\label{eq:Dirac2}
\end{eqnarray}
with

\begin{equation}
V_1 = \frac{\sqrt{\Delta} |\kappa|}{r^2}(|\kappa| \sqrt{\Delta} + \frac{r}{2}\frac{d\Delta}{dr} - \Delta),
\hspace{10pt}
\left( \kappa = \mbox{j} + \frac{1}{2}, \mbox{j = l} + \frac{1}{2} \right)
\label{eq:Potencial1}
\end{equation}

\begin{equation}
V_2 = \frac{\sqrt{\Delta} |\kappa|}{r^2}(|\kappa| \sqrt{\Delta} - \frac{r}{2}\frac{d\Delta}{dr} + \Delta),
\hspace{10pt}
\left( \kappa = - \mbox{j} - \frac{1}{2}, \mbox{j = l} - \frac{1}{2} \right).
\label{eq:Potencial2}
\end{equation}

\subsection{Quasinormal modes for the massless Dirac field}

We will evaluate the quasinormal modes using the same WKB approach used for scalar modes, and in principle we have two sets of modes. One for the function $F$ and another for the function $G$. The two potentials $V_1$ and $V_2$, however, are supersymmetric partners derived from the same superpotential, and has been show that potentials related in this way possess the same spectra of quasinormal modes for the asymptotically flat case \cite{Anderson:1991kx}. The argument has been extended to the AdS case in \cite{Giammatteo:2004wp}. In this work we calculated the frequencies using both potentials and they agree with an error less than $1\%$.

\begin{figure}
\centering
\includegraphics[scale=0.7]{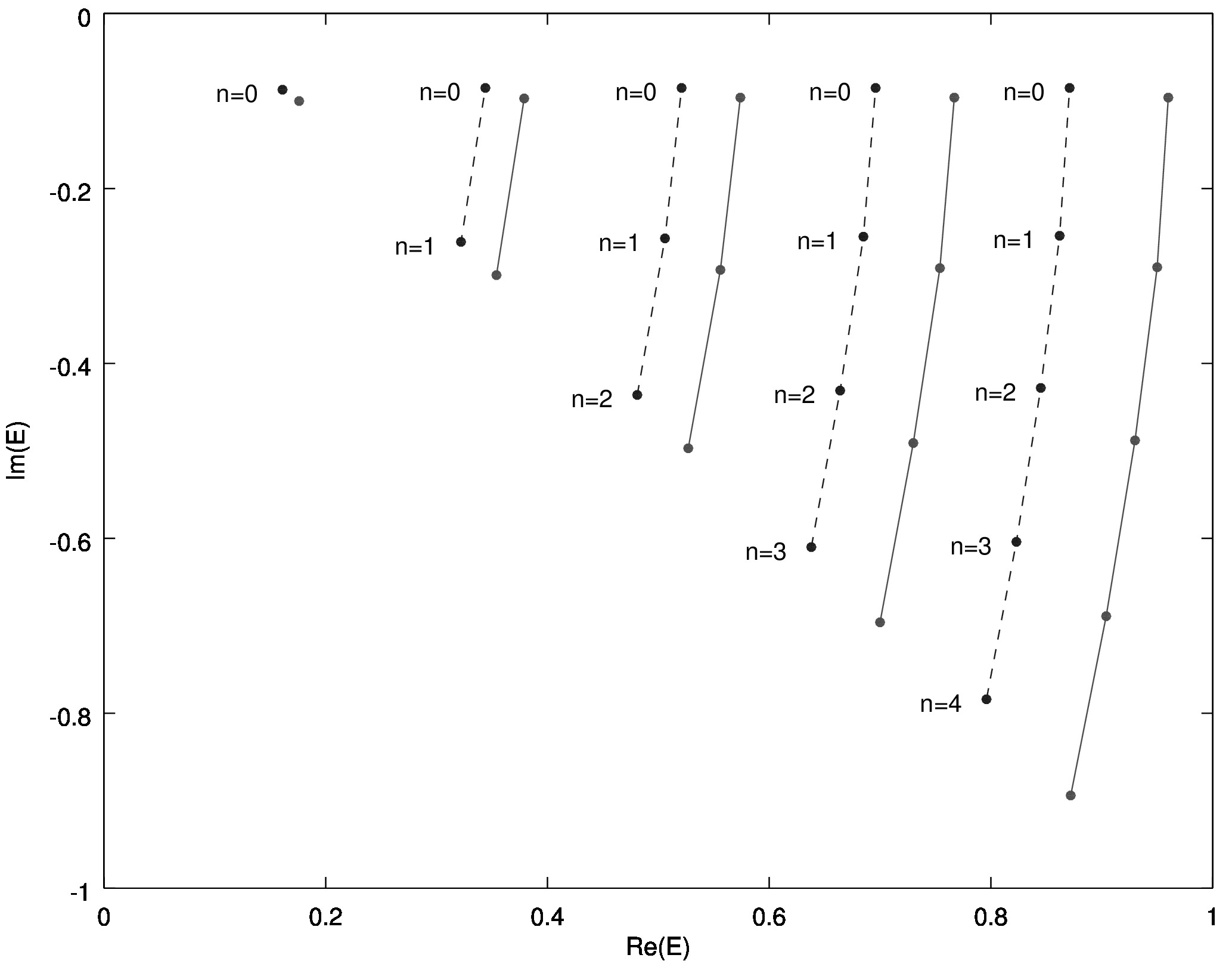}
\caption{The dotted line represents the massless Dirac quasinormal modes for a black hole with a global monopole in $f(R)$ theory, for $\psi_{0} = 0.02$ and $8 \pi \eta^2 = 10^{-5}$. The continuous line represents the same modes for the Schwarzschild solution.}
\label{fig:modesSpinHalf}
\end{figure}

Some calculated values are listed in Table \ref{tb:tableSpinHalf}, and the frequencies are plotted in Figure (\ref{fig:modesSpinHalf}). The data are compared with the values obtained for the Schwarzschild case \cite{Cho:2003qe}.

\begin{table}[]
\centering
\caption{Spin $1/2$ quasinormal modes for $\psi_0 = 0.02$ and $8 \pi \eta^2 = 10^{-5}$.}
\label{my-label}
\begin{tabular}{p{0.2\textwidth}p{0.2\textwidth}p{0.2\textwidth}p{0.2\textwidth}}
\hline
 $|k|$  & n  & Re(E) & -Im(E)  \\ \hline
 1  & 0  & 0.161 & 0.087  \\
 2  & 0  & 0.344 & 0.085  \\
 2  & 1  & 0.322 & 0.261  \\
 3  & 0  & 0.521 & 0.085  \\
 3  & 1  & 0.506 & 0.257  \\
 3  & 2  & 0.481 & 0.436  \\
 4  & 0  & 0.696 & 0.085  \\
 4  & 1  & 0.685 & 0.255  \\
 4  & 2  & 0.664 & 0.431  \\
 4  & 3  & 0.638 & 0.610  \\
 5  & 0  & 0.871 & 0.085  \\
 5  & 1  & 0.862 & 0.254  \\
 5  & 2  & 0.845 & 0.428  \\
 5  & 3  & 0.823 & 0.604  \\
 5  & 4  & 0.796 & 0.784  \\
\end{tabular}
\label{tb:tableSpinHalf}
\end{table}

%
%
\section{Conclusions}
We have evaluated the massless quasinormal modes for a black hole with a global monopole in $f(R)$ theory of gravity. We did it for the spin $0$ and spin $1/2$ cases using the third order WKB approximation. We have focused on the role played by the parameter $\psi_{0}$, where $\psi_{0} \rightarrow 0 $ means we are going from $f(R)$ gravity to Einstein's gravity. The presented metric is only valid for $\psi_{0} \ll 1$ and we have explicitly showed that as $\psi_{0}$ grows we can no longer expect a well-behaved effective potential for the scalar field, and so the proposed constraint for the $\psi_{0}$ parameter is consistent. 

In the case where $8 \pi \eta^2 \ll \psi_{0} \ll 1 $, the quasinormal mode frequencies are well defined and indicates a stable black hole solution with a global monopole inside it. We have compared our obtained results with the frequencies for the Schwarzschild black hole and found that they follow the same pattern, both for the spin $0$ and spin $1/2$ cases.

%
%
\ack J.P.M.G. was supported by CAPES Fellowship. H.S.V. and V.B.B. would like to thank Conselho Nacional de Desenvolvimento Cient\'{i}fico e Tecnol\'{o}gico (CNPq) for partial financial support.
%
%


\section*{References}
\begin{thebibliography}{99}

\bibitem{Kokkotas:1999bd} 
  K.~D.~Kokkotas and B.~G.~Schmidt,
  Living Rev.\ Rel.\  {\bf 2}, 2 (1999)
  [gr-qc/9909058].
  
\bibitem{Konoplya:2011qq} 
  R.~A.~Konoplya and A.~Zhidenko,
  Rev.\ Mod.\ Phys.\  {\bf 83}, 793 (2011)
  [arXiv:1102.4014 [gr-qc]].  

\bibitem{Schutz:1985zz} 
  B.~F.~Schutz and C.~M.~Will,
  Astrophys.\ J.\  {\bf 291}, L33 (1985).

\bibitem{Iyer:1986np} 
  S.~Iyer and C.~M.~Will,
  Phys.\ Rev.\ D {\bf 35}, 3621 (1987).

\bibitem{Nollert:1999ji} 
  H.~P.~Nollert,
  Class.\ Quant.\ Grav.\  {\bf 16}, R159 (1999).

\bibitem{Leaver:1990zz} 
  E.~W.~Leaver,
  Phys.\ Rev.\ D {\bf 41}, 2986 (1990).

\bibitem{Andersson:1996xw} 
  N.~Andersson and H.~Onozawa,
  Phys.\ Rev.\ D {\bf 54}, 7470 (1996)
  [gr-qc/9607054].
  
\bibitem{Wang:2000gsa} 
  B.~Wang, C.~Y.~Lin and E.~Abdalla,
  Phys.\ Lett.\ B {\bf 481}, 79 (2000)
  [hep-th/0003295].
  
\bibitem{Hod:2003jn} 
  S.~Hod,
  Phys.\ Rev.\ D {\bf 67}, 081501 (2003)
  [gr-qc/0301122].
  
\bibitem{Berti:2003zu} 
  E.~Berti and K.~D.~Kokkotas,
  Phys.\ Rev.\ D {\bf 68}, 044027 (2003)
  [hep-th/0303029].
  
\bibitem{Jing:2005pk} 
  J.~l.~Jing and Q.~y.~Pan,
  Nucl.\ Phys.\ B {\bf 728}, 109 (2005)
  [gr-qc/0506098].
  
\bibitem{Yoshida:2010zzb} 
  S.~Yoshida, N.~Uchikata and T.~Futamase,
  Phys.\ Rev.\ D {\bf 81}, 044005 (2010).
  
\bibitem{Yang:2012he} 
  H.~Yang, D.~A.~Nichols, F.~Zhang, A.~Zimmerman, Z.~Zhang and Y.~Chen,
  Phys.\ Rev.\ D {\bf 86}, 104006 (2012)
  [arXiv:1207.4253 [gr-qc]].
  
\bibitem{Cvetic:2013lfa} 
  M.~Cvetic and G.~W.~Gibbons,
  Phys.\ Rev.\ D {\bf 89}, no. 6, 064057 (2014)
  [arXiv:1312.2250 [gr-qc]].
  
\bibitem{Richartz:2014jla} 
  M.~Richartz and D.~Giugno,
  Phys.\ Rev.\ D {\bf 90}, no. 12, 124011 (2014)
  [arXiv:1409.7440 [gr-qc]].                

\bibitem{Oliveira:2015vqa} 
  L.~A.~Oliveira, V.~Cardoso and L.~C.~B.~Crispino,
  Phys.\ Rev.\ D {\bf 92}, no. 2, 024033 (2015).
  
\bibitem{Okuzumi:2007hf} 
  S.~Okuzumi and M.~a.~Sakagami,
  Phys.\ Rev.\ D {\bf 76}, 084027 (2007)
  [gr-qc/0703070 [GR-QC]].
    
\bibitem{Barcelo:2007ru} 
  C.~Barcelo, A.~Cano, L.~J.~Garay and G.~Jannes,
  Phys.\ Rev.\ D {\bf 75}, 084024 (2007)
  [gr-qc/0701173].      



\bibitem{Sotiriou:2008rp} 
  T.~P.~Sotiriou and V.~Faraoni,
  Rev.\ Mod.\ Phys.\  {\bf 82}, 451 (2010)
  [arXiv:0805.1726 [gr-qc]].
  
\bibitem{Nojiri:2009kx} 
  S.~Nojiri, S.~D.~Odintsov and D.~Saez-Gomez,
  Phys.\ Lett.\ B {\bf 681}, 74 (2009)  
  [arXiv:0908.1269 [hep-th]].

\bibitem{Starobinsky:1980te} 
  A.~A.~Starobinsky,
  Phys.\ Lett.\ B {\bf 91}, 99 (1980).

\bibitem{Planck:2013jfk} 
  P.~A.~R.~Ade {\it et al.} [Planck Collaboration],
  Astron.\ Astrophys.\  {\bf 571}, A22 (2014)
  [arXiv:1303.5082 [astro-ph.CO]].
  
\bibitem{'tHooft:1974qc} 
  G.~'t Hooft,
  Nucl.\ Phys.\ B {\bf 79}, 276 (1974).  
  
\bibitem{Barriola:1989hx} 
  M.~Barriola and A.~Vilenkin,
  Phys.\ Rev.\ Lett.\  {\bf 63}, 341 (1989).  

\bibitem{Barros:1997fi}
  A.~Barros and C.~Romero,
  Phys.\ Rev.\ D {\bf 56} (1997) 6688
  [gr-qc/9707040].

\bibitem{Carames:2011uu} 
  T.~R.~P.~Carames, E.~R.~Bezerra de Mello and M.~E.~X.~Guimaraes,
  Int.\ J.\ Mod.\ Phys.\ Conf.\ Ser.\  {\bf 3}, 446 (2011)
  [arXiv:1106.4033 [gr-qc]].

\bibitem{Man:2013sf} 
  J.~Man and H.~Cheng,
  Phys.\ Rev.\ D {\bf 87}, no. 4, 044002 (2013)
  [arXiv:1301.2739 [hep-th]].
    
\bibitem{Carames:2011xi} 
  T.~R.~P.~Carames, E.~R.~Bezerra de Mello and M.~E.~X.~Guimaraes,
  Mod.\ Phys.\ Lett.\ A {\bf 27}, 1250177 (2012)
  [arXiv:1111.1856 [gr-qc]].    
    
\bibitem{Iyer:1986nq} 
  S.~Iyer,
  Phys.\ Rev.\ D {\bf 35}, 3632 (1987).  

\bibitem{Konoplya:2003ii} 
  R.~A.~Konoplya,
  Phys.\ Rev.\ D {\bf 68}, 024018 (2003)
  [gr-qc/0303052].
  
\bibitem{Anderson:1991kx} 
  A.~Anderson and R.~H.~Price,
  Phys.\ Rev.\ D {\bf 43}, 3147 (1991).  

\bibitem{Giammatteo:2004wp} 
  M.~Giammatteo and J.~l.~Jing,
  Phys.\ Rev.\ D {\bf 71}, 024007 (2005)
  [gr-qc/0403030].

\bibitem{Cho:2003qe} 
  H.~T.~Cho,
  Phys.\ Rev.\ D {\bf 68}, 024003 (2003)
  [gr-qc/0303078].

%
%
\end{thebibliography}
\end{document}